\documentclass[showpacs,twocolumn,aps,preprintnumbers,letterpaper]{revtex4}
\usepackage{amsmath,amssymb}
\usepackage{epsfig}
\usepackage{graphicx}
\usepackage{amsmath}
\usepackage{slashed}
\usepackage{amsfonts}
\usepackage{epstopdf}
\newcommand{\be}{\begin{equation}}
\newcommand{\ee}{\end{equation}}
\newcommand{\bear}{\begin{eqnarray}}
\newcommand{\eear}{\end{eqnarray}}
\newcommand{\ba}{\begin{array}}
\newcommand{\ea}{\end{array}}

\def\be{\begin{eqnarray}}
\def\ee{\end{eqnarray}}
\def\bea{\be}
\def\eea{\ee}

\def\roughly#1{\mathrel{\raise.3ex\hbox{$#1$\kern-.75em%
\lower1ex\hbox{$\sim$}}}}

\begin{document}

\title{Hydrodynamics of the Polyakov Line in SU$(N_c)$ Yang-Mills}

\author{Yizhuang Liu}
\email{yizhuang.liu@stonybrook.edu}
\affiliation{Department of Physics and Astronomy, Stony Brook University, Stony Brook, New York 11794-3800, USA}

\author{Piotr Warcho\l{}}
\email{piotr.warchol@uj.edu.pl}
\affiliation{M. Smoluchowski Institute of Physics, Jagiellonian University, PL-30348 Krakow, Poland}

\author{Ismail Zahed}
\email{ismail.zahed@stonybrook.edu}
\affiliation{{}Department of Physics and Astronomy, Stony Brook University, Stony Brook, New York 11794-3800, USA}

%%%%%%%%%

\date{\today}
\begin{abstract}
We discuss a hydrodynamical description of the eigenvalues of the Polyakov line at large but finite $N_c$
for Yang-Mills theory in even and odd space-time dimensions. The 
hydro-static solutions for the eigenvalue densities are shown to interpolate between a uniform  distribution in the confined
phase and a localized distribution in the de-confined phase. The resulting critical temperatures are in overall agreement 
with those measured  on the lattice over a broad range of $N_c$, and are consistent with the string model results at $N_c=\infty$. 
The stochastic relaxation of the eigenvalues of the Polyakov line
out of equilibrium is captured by a hydrodynamical instanton. An estimate of the probability of formation of a Z(N$_c)$
bubble using  a piece-wise sound wave is suggested. 
\end{abstract}

%\pacs{11.25.Tq, 13.60.Hb,13.85.Lg}%11.25.Tq 	Gauge/string duality  13.60.Hb for deep-inelastic structure functions; 13.85.Lg 	Total cross sections

%\pacs{11.25.Tq, 13.60.Hb,13.85.Lg}
%11.25.Tq 	Gauge/string duality  13.60.Hb for deep-inelastic structure functions; 13.85.Lg 	Total cross sections

%11.15.Kc	Classical and semiclassical techniques
%11.30.Rd	Chiral symmetries
%12.38.Lg	Other nonperturbative calculations

%11.15.Ex	Spontaneous breaking of gauge symmetries

%12.38.Aw	General properties of QCD (dynamics, confinement, etc.)
%11.10.Wx	Finite-temperature field theory
%12.38.Mh	Quark-gluon plasma
%71.10.Pm	Fermions in reduced dimensions (anyons, composite fermions, Luttinger liquid, etc.) (for anyon mechanism in superconductors, see 74.20.Mn)
 %73.43.Lp	Collective excitations
%11.15.Pg	Expansions for large numbers of components (e.g., 1/Nc expansions)

\maketitle

\setcounter{footnote}{0}

%\baselineskip 18pt \pagebreak
%\renewcommand{\thepage}{\arabic{page}}
%\tableofcontents
%\pagebreak

%\section{Introduction}

{\bf 1. Introduction.\,\,} Lattice simulations of Yang-Mills theory in even and odd dimensions show that the confined
phase is center symmetric~\cite{LATTICE,LATTICE3}. At high temperature Yang-Mills theory is in a deconfined phase
with broken center symmetry. The transition from a center symmetric to a center broken phase
is non-perturbative and is the topic of intense numerical and effective model calculations~\cite{OGILVIE}
(and references therein).  Of particular interest are the semi-classical descriptions  and matrix models. 

In the semi-classical approximations, the confinement-deconfinement transition is understood
as the breaking of  instantons into a dense plasma of dyons  in the confined phase and their
re-assembly into instanton molecules in the deconfined phase~\cite{DP,US}. A mechanism 
similar to the Berezinsky-Kosterlitz-Thouless  transition in lower dimensions~\cite{BKT}, and to the transition 
from insulators to superconductors in topological materials~\cite{SUDHIP}. 
In matrix models, the Yang-Mills theory is simplified to the eigenvalues of the Polyakov line 
and an effective potential is used with  parameters fitted to the bulk pressure to study such a
transition~\cite{ROB,KEVIN}, in the spirit of the strong coupling transition in the Gross-Witten model~\cite{GROSS}.

Matrix models for the Polyakov line share much in common with unitary matrix models in the
general context of random matrix theory~\cite{MEHTA}. The canonical example is Dyson circular unitary
ensemble and its analysis in terms of orthogonal polynomials or a one-component Coulomb plasma. 
The Dyson circular unitary ensemble relates to the  Calogero-Sutherland model
which is an effective model for quantum Luttinger liquids. A useful analysis of this model uses the collective 
quantization method developed in~\cite{JEVICKI} with its hydrodynamical 
interpretation~\cite{ALEXIOS,SASHA}.

In this letter we  develop a hydrodynamical description of the gauge invariant eigenvalues
of the Polyakov line for an SU(N$_c$) Yang-Mills theory at large but finite $N_c$.  
We will use it to derive the following new results: 1/ 
a hydrostatic solution for the eigenvalue density that interpolates between a confining (uniform) and de-confining
(localized) phase; 2/ explicit critical temperatures  for the Yang-Mills transitions in $1+2$ and $1+3$ dimensions;
3/ a hydrodynamical instanton for the density distribution that captures the stochastic relaxation of the eigenvalues of
the Polyakov line; 4/ an estimate of the fugacity or probability to form  a Z(N$_c$) bubble using a piece-wise sound-wave.

%Dyson diffusion equation for the unitary random matrix models as noted recently in the context
%of Wilson loops in $1+1$ dimensions~\cite{MACIEK}.

\vskip 0.5cm

{\bf 2. Polyakov line in $1+2$ dimensions.\,\,}
The matrix model partition function for the eigenvalues of the Polyakov line
for SU($N_c$) in $1+2$ dimensions was discussed in~\cite{ROB}. If we denote by 
${\rm diag}(e^{i\theta_1}, ..., e^{i\theta_{N_c}})$ with $\sum_{i}\theta_{i}=0$ the gauge invariant
eigenvalues of the Polyakov line, then~\cite{ROB} 

\be
Z[\alpha, \beta]=\int \prod_{i=1}^{N_c} d\theta_i\prod^{N_c}_{i<j}|z_{ij}|^{\beta(T)} e^{-\alpha(T)\sum_{i<j}V(|z_{ij}|)}
\label{1}
\ee
with $z_{ij}=z_i-z_j$ and $z_i=e^{i\theta_i}$. The perturbative
potential $V(z_{ij})$ is center
symmetric and quadratic in leading order or $V(|z_{ij}|)\approx |z_{ij}|^2$, with
 $\alpha(T)= T^2\,V_2/2\pi$ and $V_2$ the spatial 2-volume~\cite{ROB}. 
%The Van-dermond contribution 
%is non-perturbative and extensive with $V_2$. 
The  mass expansion of the one-loop determinant 
gives $ \beta(T)=m_D^2\,V_2/\pi$~\cite{ROB}. The Debye mass is self-consistently defined as
$m_D^2=N_cg^2T({\rm ln} (T/m_D)+C)/2\pi$~\cite{DHOKER} to tame all 
infra-red divergences, with $C\approx 1.3$ from lattice simulations~\cite{BENGT,USDATA}.

(\ref{1}) can be regarded as the normalization of the squared and
real many-body wave-function ${\bf \Psi}_0[z_i]$
%\be
%Z[\alpha, \beta]=\int \prod_{i=1}^{N_c} d\theta_i\,\left|\Psi_0[z_i]\right|^2
%\label{3}
%\ee
%\be
%\Psi_0[z_i]=\prod^{N_c}_{i<j}|z_i-z_j|^{\frac {\beta(T)} 2} e^{-\frac {\alpha(T)}2 \sum_{i<j}V(|z_i-z_j|)}\equiv e^{-S[z]}
%\label{3}
%\ee
which  is the zero-mode solution to the Shrodinger equation $H_0\Psi_0=0$ with
the self-adjoint squared Hamiltonian

\be
H_0\equiv\sum_{i=1}^{N_c}\left(-\partial_i+{\bf a}_i\right)\left(\partial_i+{\bf a}_i\right)
\label{4}
\ee
with $\partial_i\equiv \partial/\partial\theta_i$ and the pure
gauge potential ${\bf a}_i\equiv \partial_iS$. Here $S[z]=-{\rm ln}\Psi_0[z]$
is half the energy  in the defining partition function in (\ref{1}). In (\ref{4}) the mass parameter is $1/2$.

\vskip 0.5cm
{\bf 3. Hydrodynamics.\,\,}
We can use the collective coordinate method in~\cite{JEVICKI}  to re-write (\ref{4}) in terms of the
density of eigenvalues as a collective variable $\rho(\theta) =\sum^{N_c}_{i=1}\delta(\theta-\theta_i)$. 
For that,  we re-define $H_0\rightarrow H$ through a similarity transformation to 
re-absorb the diverging 2-body part induced by the 
Vandermond contribution  $\Delta=\prod_{i<j}|z_{ij}|^{\beta(T)}$,  i.e. $\Psi=\Psi_0/\sqrt{\Delta}$  and 
$\sqrt{\Delta}\,H=H_0\sqrt{\Delta}$.
%\be
%H\Psi=\sum^{N_c}_{i=1}\frac{1}{\Delta^{\beta(T)}}
%\left(-\partial_i+a_i\right)\Delta^{\beta(T)}\left(\partial_i+a_i\right)\Psi
%\label{5X1}
%\ee
%with $2a_i=\alpha(T)\partial_i\sum_{j<k}V|z_{jk}|$.
%\be
%A_i=\frac 12 \alpha(T) \sum_{i<j}\partial_iV(|z_i-z_j|)
%\label{5X2}
%\ee
Now $H$ is of the general form discussed in~\cite{JEVICKI} and is  amenable
after some algebra to 

\be
H=\int d\theta \left(\partial_{\theta}\pi\rho\, \partial_{\theta}\pi+ \rho {\bf u}[\rho]\right)
\label{5X3}
\ee
with the potential-like contribution

\be
{\bf u}[\rho]
%\frac{1}{4}\frac{(\partial-\theta \rho-\pi\beta(T)\rho \rho_H)^2}{\rho}-\pi \beta(T) A \rho \rho_H-\rho \partial_{\theta} A+\rho A^2\nonumber \\
=\left(A(\theta)-\frac{\pi\beta(T)\rho_H}{2}+\frac 12 {\partial_\theta}{\rm ln}\rho\right)^2\equiv {\bf A}^2
\label{5X4}
\ee
Here  

\be
A(\theta)=\frac 12 \alpha(T)\int  d\theta^\prime\rho(\theta^\prime) \,\partial_\theta V\left(2\,{\rm sin}\left(\frac{\theta-\theta^\prime}2\right)\right)
\ee
and $\rho_H$ is the periodic Hilbert transform of $\rho$

\be
[\rho]_H\equiv \rho_H(\theta)=\frac 12 \frac {\rm P}\pi \int {\rho(\theta^\prime)}\,{\rm cotan}\left(\frac {\theta-\theta^\prime}2 \right)
\label{7}
\ee
As conjugate pairs, $\pi(\theta)$ and  $\rho(\theta)$  satisfy 
the equal-time commutation rule
$[\pi(\theta),\rho(\theta^\prime)]=-i\left(\delta(\theta-\theta^\prime)-{1}/{2\pi}\right)$.
%Since $[\partial_\theta\pi(\theta),{\bf A}(\theta)]$ is ultra-local, it can be dropped in the re-writing
%of (\ref{5X3}). 
We identify the collective fluid velocity with $v=\partial_\theta \pi$
and re-write (\ref{5X3}) in the more familiar hydrodynamical form

\be
H\approx \int d\theta \rho(\theta) \left( v^2+ {\bf u}[\rho]\right)\approx \int d\theta \rho(\theta) \left|v+i{\bf A}\right|^2
\label{7X1}
\ee
modulo ultra-local terms. 
%The first contribution in (\ref{7X1}) is the fluid kinetic energy,
%while the second contribution is the fluid potential energy.  
The Heisenberg equation for $\rho$ yields the current conservation law 
$\partial_t\rho=-2\partial_\theta \left(\rho  v\right)$, and the Heisenberg equation 
for $v$ gives the  Euler equation

\bea
\label{10}
%&&\partial_t\rho=i[H, \rho(t, \theta)]=-2\partial_\theta \left(\rho  v\right)\\
&&\partial_t v=i[H, v]=\\
&&-\partial_\theta\left(v^2+{\bf A}^2-\partial_\theta {\bf A}
-{\bf A}{\partial_\theta{\rm ln}\rho}+\pi \beta[{\bf A}\rho]_H-2\alpha[{\bf A}\rho]_S\right)\nonumber
\eea
with the sine-transform $[{\bf A}\rho]_S=\int {\rm sin}(\theta-\theta'){\bf A}(\theta')\rho(\theta')$. 
Note that all the relations  hold for large but finite $N_c$.

\vskip 0.5cm

{\bf 4. Hydro-static solution.\,\,}
The static  hydrodynamical density follows from the minimum of (\ref{7}) with  $v(\theta)=0$,

\be
\beta(T)\pi\rho_H(\theta)-\partial_\theta{\rm ln}\rho(\theta)= 2A(\theta)
%2\alpha\omega(\theta)
\label{12}
\ee
To solve (\ref{12}), we insert the leading quadratic contribution $A(\theta)\approx 2\alpha(T){\rm sin}^2(\theta/2)$
in (\ref{12})
%which follows from the one-loop effective potential in any dimension. With this in mind, (\ref{12}) now reads

\be
\rho\rho_H-a \partial_\theta \rho=bc_1\rho\,{\rm sin} (\theta)
\label{16X}
\ee
with $a\equiv {1}/{\pi\beta(T)}$, $b\equiv {2\alpha(T)}/{\beta(T)}$ and $c_1$ the first moment of the density or
$\pi c_1\equiv  \int^{2\pi}_0 \rho(\theta){\rm cos}\theta  d\theta$.
Let $\rho_0=N_c/2\pi$ be the uniform eigenvalue density and $\rho_1=\rho-\rho_0$ its deviation. 
Consider the Cauchy transform 

\be
G(z)=\frac{1}{\pi i}\int_{\cal C} \frac{\rho_1(\eta)}{\eta-z}d\eta
\label{16X2}
\ee
with $\eta=e^{{i\theta}}$. The contour ${\cal C}$ is counter-clockwise along the unit circle. $G(z)$ is a
holomorphic function in the complex z-plane. Let $G^+$ and $G^-$ be its realization inside and outside  ${\cal C}$
respectively, so that 

\bea
G^{\pm}(z\rightarrow e^{i\theta})=\pm \rho_1(\theta)+i\rho_H(\theta)
%\nonumber\\
%&&G^{-}(z\rightarrow e^{i\theta})=-\rho_1(\theta)+i\rho_H(\theta)
\label{16X3}
\eea
%Thus, $\rho_1(\theta)+i\rho_H(\theta)$ is the boundary value of the function $G^{+}(z)$ defined inside the circle $C$. 
We now carry the Hilbert transform on both sides of (\ref{16X}).
Setting $G(z)=G^+(z)$ and using $2[\rho_1\rho_H]_H=\rho_H^2-\rho_1^2$  , we have for (\ref{16X})

\be
\frac 12 {G^2}+(\rho_0-\frac 12 {bc_1(z-z^{-1})})G+az\partial_zG=bc_1\rho_0z+\frac 12 {bc_1^2}\nonumber\\
\label{16X5}
\ee
on the boundary ${\cal C}$, thus within the circle. Here, we should require $G(z=0)=0$ to ensure that $\rho_1$ integrates to zero.
%In the general case and for  non-vanishing $a$, (\ref{16X5}) can be transformed to an ordinary differential equation using the change of variable
%$G(z)=2az\partial_z{\rm ln}u(z)$. 

\begin{figure}[htb]
\begin{center}
\includegraphics[width=7cm]{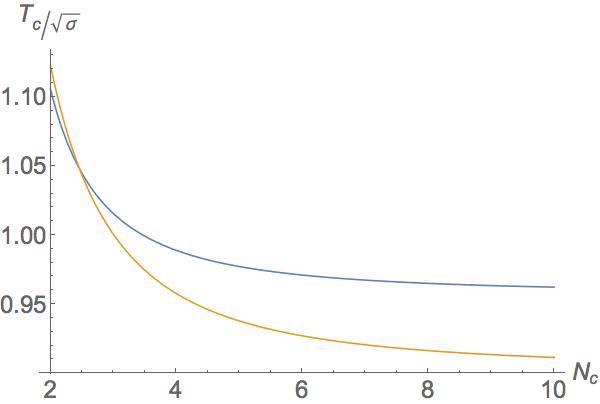}
\caption{$T_c/\sqrt{\sigma_1}$ versus  $N_c$ in  (\ref{18Z2}) (upper curve) compared to the numerical fit  to the lattice results 
(lower curve) from~\cite{TEPER2}. }
\label{TC2-data}
\end{center}
\end{figure}

$a\approx 1/V_2$  is subleading and will be dropped. Thus (\ref{16X5}) is algebraic in $G(z)$.
Since $\rho(\theta) =\rho_0+{\rm Re}\,G^+(z=e^{i\theta})$, 
careful considerations of the singularity structures of the quadratic solutions to (\ref{16X5}) 
yield ($\Theta$ is a step function)

\be
\rho(\theta)=\sqrt{bc_1}
({\rm cos}\theta+1)^{\frac 12}({\rm cos}\theta-{\rm cos}\theta_0)^{\frac 12}\, \Theta(|\theta_0|-|\theta|)\nonumber\\
\label{16X7}
%\\\rho_0^2+bc_1^2-2\rho_0bc_1=0,{\rm cos}\theta_0=1-\frac{2\rho_0}{bc_1}\\\frac{c_1}{\rho_0}=1+
%\sqrt{\Delta(\theta)}, \\\Delta(\theta)=
%\sqrt{1-\frac{1}{b}}
\ee
The analytic properties of $G(z)$ fix 
$c_1/\rho_0=1+(1-1/b)^{\frac 12}$ and 
%\be
%\frac{c_1}{\rho_0}=1+ \sqrt{1-\frac{1}{b}}
%\label{16X8}
%\ee
 $\theta_0$ at ${\rm cos}\,\theta_0=1-{2\rho_0}/{bc_1}$. For $b<1$ the non-uniform
solution with $\rho_1\neq 0$ is absent.  For $b\gg 1$,  $c_1\rightarrow 2\rho_0$ and

\be
\rho(\theta)\rightarrow \frac{N_c}{2\pi}\sqrt{8b-4b^2\theta^2}
\label{18Z1}
\ee
%which is Wigner semi-circular distribution with $b={2\alpha(T)}/{\beta(T)}$.
Therefore (\ref{16X7}) interpolates between a uniform density distribution $\rho_0$ (confined phase)
and a Wigner semi-circle (deconfined phase)  with a transition at  
$b=1$ or $T_c=m_D$. In $1+2$ dimensions the fundamental string tension is given to a good accuracy by
$\sqrt{\sigma_1}/g^2N_c=((1-1/N_c^2)/8\pi)^{\frac 12}$~\cite{NAIR}. Thus the  ratio in $1+2$ dimensions

\be
\frac{T_c}{\sqrt{\sigma_1}}=\frac C{2\pi}\,\left(\frac {8\pi}{1-1/N_c^2}\right)^{\frac 12}\rightarrow \sqrt{\frac{2}\pi}\,C
\label{18Z2}
\ee
%The rightmost result is the large $N_c$ limit of $1.0161$.
with $C\approx 1.3$~\cite{BENGT,USDATA}. In Fig.~\ref{TC2-data} we show the behavior of (\ref{18Z2}) (upper curve) versus $N_c$,
in comparison to the numerical  fit $T_c/\sqrt{\sigma_1}=0.9026+0.880/N_c^2$
to the lattice results (lower curve)  in~\cite{TEPER2}.  Amusingly, (\ref{18Z2}) at large $N_c$
is consistent with $\sqrt{3/\pi}$ in the string model~\cite{ALVAREZ}.

\vskip .5cm
{\bf 5. Dyson Coulomb gas.\,\,}
We note that (\ref{12}) coincides with the saddle point equation to (\ref{1}) by re-writing
it using Dyson charged particle analogy on $S^1$ with the energy
$2S[z]=\sum_{i<j}G(z_{ij})$
%\be
%2S[z]=2{\rm ln}|\Psi_0[z]|=\sum_{i<j}^{N_c}G(z_{ij})
%=\frac{1}{2}\rho {\bf G} \rho-\frac{1}{2}{\bf G}_R\rho
%\label{D2}
%\%ee
and the pair interaction

\be
{G}(z_{ij})=-{\rm ln}|z_{ij}|^{\beta(T)}+{\alpha (T)} V( |z_{ij}|)\equiv {\bf G}(\theta_i-\theta_j)
\label{D3}
\ee
%${\bf G}_R$ regulates the divergence in the 2-particle kernel
%\be
%G_R(\theta)={\lim_{\theta\to\theta^\prime}}\left(G(\theta ,\theta^\prime)+{G}(\rho |\theta-\theta^\prime|\right)
%\label{D6}
%\ee
%On $S^1$ we identify $L_{ij}=|z_{ij}|$ with the chordal length and $l(\rho)\approx 1/\rho$ 
%as the distance of closest encounter which will be used as a short distance cutoff or core. 
%This physical subtraction is consistent with the subtraction through  Hilbert transforms in integral forms.
At large $N_c$ the ensemble described by (\ref{1}) is sufficiently dense  to allow the change in the measure. 
%\be
%\prod_{i}d\theta_i\approx e^{{\cal S}[\rho ]} D\rho
%\label{D7}
%\ee
%with ${\cal S}[\rho]=\int d\theta\rho(\theta){\rm ln}(\rho_0/\rho(\theta))$ the Boltzmann entropy~\cite{MEHTA}.
%\be
%{\cal S}[\rho]=-\int  d\theta\rho (\theta)\,{\rm ln}\left( \frac{\rho}{e\rho_0}\right)
%\label{D8}
%\ee
%Inserting (\ref{D2}-\ref{D7}) into (\ref{1}) yields
Following Dyson~\cite{MEHTA} we obtain

\be
Z[\alpha, \beta]\rightarrow \int D\rho \,e^{-\Gamma[\alpha, \beta; \rho]}
\label{D9}
\ee
with the effective action

\bea
\Gamma[\alpha, \beta; \rho]=&&\frac{1}{2}
\int \rho(\theta) {\bf G}(\theta-\theta^\prime) \rho(\theta^\prime) \nonumber\\
&&-\left(\frac{\beta(T)}{2}-1\right)\int d\theta \rho(\theta) {\rm ln}\left(\frac{\rho(\theta)}{\rho_0}\right) 
%\nonumber \\
%=&&\frac{1}{2}\phi{\bf A}\phi-\left(\frac{\beta}{2}-1\right)\rho \frac{\rho}{e\rho_0}
\label{D10}
\ee
The $\beta$ contribution is the self Coulomb subtraction and is consistent with the subtraction in the Hilbert transform.
The saddle point equation $\delta\Gamma/\delta\rho=0$ following from  (\ref{D9}-\ref{D10}) is in agreement with
the hydro-static equation (\ref{12}), 

\be
\frac{d}{d\theta}\frac{\delta \Gamma[\alpha, \beta; \rho]}{\delta \rho(\theta)}= 2{\bf A}=0
%\frac{\partial \rho}{\rho}-\beta \pi \rho_H+2\alpha \int {\rm sin}(\theta-\theta')\rho(\theta')\nonumber \\=2A_{eff}(\theta)
\ee

\vskip 0.5cm

{\bf 6. Hydrodynamical instanton.\,\,}
The fixed time  zero energy solution to (\ref{7X1}) is an instanton with imaginary velocity
$v=-i {\bf A}$. We have checked that this is a solution to (\ref{10}) for all times. 
The current  $j\equiv \rho v=-i\rho {\bf A}$ is conserved.Thus
$\partial_\tau\rho-2\partial_\theta (\rho {\bf A})=0$ or

\be
\partial_\tau\rho+{\beta(T)}\partial_\theta(\pi \rho\rho_H)=\partial_\theta^2\rho+2\partial_\theta(\rho{  A}(\theta))
\label{15XX}
\ee
for Euclidean times $\tau=it$. For $A=0$ and $\beta(T)=2$,   (\ref{15XX}) agrees with the viscid Burger$^\prime$s equation describing
large Wilson loops in $1+1$ dimensions~\cite{MACIEK}. Following~\cite{MEHTA}
we identify $\tau$ with the stochastic (Langevin) time.
%Dyson ensemble of random unitary matrices~\cite{MEHTA}.
(\ref{15XX}) describes the stochastic relaxation of the  eigenvalue density of the Polyakov line (out of equilibrium) 
to its asymptotic (in equilibrium) hydro-static solution.

\vskip 0.5cm
{\bf 7. Sound waves.\,\,}
The hydrodynamical action follows from standard procedure.The 
momentum $\pi(\theta)=(1/\partial_\theta)v$ is canonically conjugate to the density $\rho$,  
and the Lagrange density is ${\bf L}=\pi\partial_t\rho-H$. Thus the action
%Using the current conservation law (\ref{10}) allows for the half mass-shell action
%${\bf S}=\int dt d\theta\,{\bf L}$.
${\bf S}=\int dt d\theta\,\rho(\theta)\left(v^2-{\bf u}[\rho]\right)$, which is linearized by
%\be
%{\bf S}=\int dt d\theta\,\rho(\theta)\left(v^2-{\bf u}[\rho]\right)
%\label{15XX1}
%\ee
%The first contribution is the  fluid kinetic energy and the second contribution
%is the standard fluid potential energy.

\bea
\rho\approx\rho_0(\theta)+2\partial_{\theta}\varphi \qquad{\rm and}\qquad
\rho v\approx- \partial_t\varphi
\label{15XX3}
\eea
%which are seen to satisfy the current conservation law in leading order. 
Inserting (\ref{15XX3}) into ${\bf S}$ yields 
%the leading quadratic
%contribution

\be
{\bf S}_2=\int dt \frac {d\theta}{\rho_0(\theta)}  \, 
\left({(\partial_t \varphi)^2}-{\rho^2_0(\theta)}{W}^2[\varphi]\right)
%\left({(\partial_t \varphi)}-{\rho_0(\theta)}{W}[\varphi]\right)\nonumber\\
\label{15Z11}
\ee
with the potential

\be
W[\varphi]=2\alpha(T) [\partial_\theta\varphi]_S
%\int d\theta'{\rm sin}(\theta-\theta')\partial_{\theta'}\varphi(\theta')
-\pi \beta(T)[\partial_{\theta}\varphi]_H+
\partial_\theta\left(\frac{\partial_\theta{\varphi}}{\rho_0(\theta)}\right)
%\frac{\partial_\theta^2 \varphi}{\rho_0(\theta)}-\frac{\partial_\theta \rho_0}{\rho_0^2}\partial_\theta\varphi
\label{15XX4}
\ee
For constant $\rho_0$ and large  $N_c$, (\ref{15Z11}) simplifies to
\bea
{\bf S}_2\approx  m_D^2V_2\int dt d\theta\left({(\partial_t \varphi)^2}-(\partial_{\theta}\varphi)^2\right)
%\nonumber \\
%&&-4\rho_0\pi^3\alpha(T)(\alpha(T)-{\beta(T)}/2)\int  dt (c_1^2+b_1^2)
\label{15XX5}
\eea
%with the first harmonics 
%$\varphi=c_1(t){\rm cos}(\theta)+b_1(t){\rm sin}(\theta)+..$. (\ref{15XX5}) describes massless hydrodynamical sound modes 
after the rescaling $v_st\rightarrow t$ with  $v_s=\pi\rho_0\beta(T)$. (\ref{15XX5}) describes  sound waves in the 
large $N_c$ space of holonomies.

\vskip 0.5cm
{\bf 8. Z(N$_c$) bubble.\,\,}
In a de-confined phase of infinite volume, the Yang-Mills ground state 
settles in one of the degenerate Z(N$_c$) vacua. In a finite volume,  bubbles of different vacua may form~\cite{KEJO}.  
Consider a de-confined bubble of volume $\mathbb{V}_2$ immersed in a confined volume $\overline{\mathbb{V}}_2$.
In $\mathbb{V}_2$ all the eigenvalues are localized initially within a small $\Delta\theta$ around the origin with
$\rho(\tau=0, \theta)=N_c/\Delta\theta\equiv \rho_B$, and zero otherwise. 
%This is a piece-wise initial wave that relaxes to the uniform  and confined phase 
%over a time $\Delta\tau\approx \Delta\theta$ set by the speed of sound (here 1 after re-scaling).

Using this piece-wise wave as an initial condition we solve (\ref{15XX}) with $A=0$ for simplicity. 
For large times $\tau$, the result is

\be
\rho(\tau, \theta)\approx \rho_0-\left(\frac 2\pi \rho_B\,{\rm sin}\left(\frac{\Delta\theta}2\right)\right)\,{\rm cos}\,\theta\, e^{-v_s\tau}
\label{15YY1}
\ee
which shows the relaxation of the piece-wise wave over a time $\tau\approx 1/v_s$ set by the speed of sound.
Using  (\ref{15YY1}) in ${\bf S}$ yields the Euclidean action estimate for small $\Delta\theta$

\be
{\bf S}_{E}(\mathbb{V}_2)
%=m_D^2\mathbb{V}_2\int d\tau d\theta \, |\nabla\varphi|^2\approx 
\approx \mathbb{V}_2\left(\pi m_D\rho_B \,{\rm sin}\left(\frac{\Delta\theta}2\right)\right)^2\rightarrow \mathbb{V}_2
\left(\frac{\pi}2 N_cm_D\right)^2
%\frac 12 m_D^2\mathbb{V}_2 \left(\rho_0 \Delta\theta\right)^2
\label{15Z1}
\ee
The bubble formation probability or fugacity is $e^{-{\bf S}_{E}(\mathbb{V}_2)}$.

\begin{figure}[htb]
\begin{center}
\includegraphics[width=7cm]{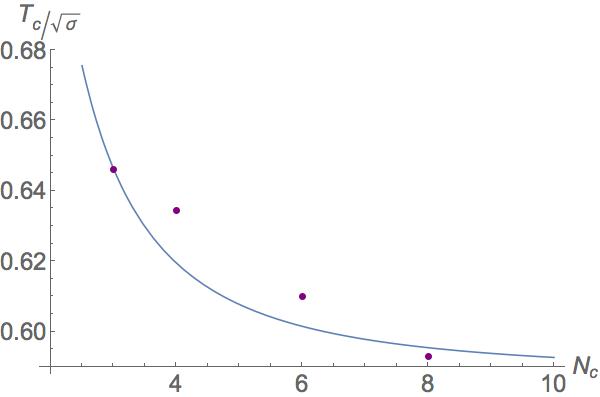}
\caption{$T_c/\sqrt{\sigma_1}$ versus $N_c$ in (\ref{26}). The dots are the lattice results from~\cite{TEPER}.}
\label{TC-data}
\end{center}
\end{figure}

\vskip 0.5cm
{\bf 9. Polyakov line  in $1+3$ dimensions.\,\,}
To extend our analysis to $1+3$ dimensions, we approximate the Yang-Mills thermal
state by a dense plasma of dyons and anti-dyons~\cite{DP,US}. This semi-classical
description reproduces a number of key features of the Yang-Mills phase both in the
confined (center-symmetric) and de-confined (center-broken) phase. There are two
key differences with the $1+2$ dimensional partition function in (1). First the many-body
energy  $2S[z]=-2{\rm ln}\Psi_0[z]$ in (1) is now shifted

\be
2S[z]\rightarrow 2S[z]-{\gamma(T)} \prod_{i}^{N_c}(\theta_{i+1}-\theta_i)^{\frac 1{N_c}}
\label{16}
\ee
with $\gamma(T)=4\pi N_c fV_3 $ and $f=4\pi\Lambda^4/Tg^4$ the dyon fugacity~\cite{DP}.
Second and more importantly $\beta(T)=2$ and is not extensive with  the spatial
3-volume $V_3$. Finally, $\alpha(T)=T^3V_3/3$. Since
%\be
%\partial_i\sum_{j=1}^{N_c}{\rm ln}\nu_j=\frac{1}{\theta_i-\theta_{i+1}}+\frac{1}{\theta_i-\theta_{i-i}}\approx-\partial_\theta {\rm ln}\rho (\theta_i)
%\label{17}
%\ee
$(\theta_{i+1}-\theta_i)\approx 1/2\pi\rho(\theta_i)$, then in the continuum the additional string
of factors in (\ref{16}) is 

\be
\prod_{i}^{N_c}(\theta_{i+1}-\theta_i)^{\frac 1{N_c}}\rightarrow e^{\frac 1{N_c}\int d\theta\rho(\theta){\rm ln}(1/2\pi\rho(\theta))}
%\equiv \frac{2\pi}{N_c}
\label{18}
\ee
With this in mind, a re-run of the preceding arguments yields the Hamiltonian in (\ref{5X3}-\ref{5X4}) with the shifted potential

\be
A\rightarrow A+\frac{\gamma(T)}{4\pi N_c^2}e^{-\gamma_0[\rho]}\partial_\theta{\rm ln}\rho(\theta)
%e^{-\frac 1{N_c}\int d\theta  \rho(\theta) {\rm ln}(\rho(\theta)/N_c)}\partial_\theta{\rm ln}\rho(\theta)
\ee
and  $N_c{\rm ln}\gamma_0[\rho]=\int d\theta  \rho(\theta) {\rm ln}(\rho(\theta)/N_c)$.
The hydro-static equation (\ref{12}) now reads

\bea
\beta\pi\rho_H(\theta)-2A(\theta)
=\left(1+\frac{\gamma(T)}{4\pi N_c^2}e^{-\gamma_0[\rho]}\right)\partial_\theta{\rm ln}\rho(\theta)
\label{20}
\eea
The $\beta=2$ contribution is now sub-leading and can be dropped. The corresponding solution to (\ref{20}) 
is  a localized density for $\pi c_1=\int_0^{2\pi}d\theta\rho(\theta){\rm cos}\theta\neq 0$, and a
uniform density $\rho_0=N_c/2\pi$  for $c_1=0$. 
%\be
%\rho(\theta)={\bf C}\,e^{\frac{8\pi^2 N^2\alpha\gamma_0}{\gamma}{c}_1{\rm cos}\theta}
%\,\,\,{\rm and}\,\,\,
%{\bf c}_1\equiv \int d\theta\, \rho \,{\rm cos}(\theta)
%,\int d\theta \rho(\theta)=N\\
%\label{21}
%\ee
%\be
%\gamma_0=e^{\frac 1{N_c}\int d\theta  \rho(\theta) {\rm ln}(\rho(\theta)/N_c)}
%\label{22}
%\ee
Specifically

\be
\frac{{\rho}(\theta)}{\rho_0}=\frac{e^{\frac{8\pi\alpha\gamma_0}{\gamma'}c^\prime{\rm cos}\theta}}{ I_0(\frac{8\pi\alpha\gamma_0}{\gamma'}c^\prime)}
\label{23}
\ee
with $c^\prime={c}_1/N_c$ and $\gamma^\prime=\gamma/N_c^3$. The two parameters
$\eta=8\pi\alpha(T)/\gamma^\prime$ and $x=c^\prime\,\eta\gamma_0$ are fixed by
the transcendental equations

\bea
\frac{I_1(x)}{I_0(x)}=\frac{\pi x}{\eta\gamma_0}\qquad {\rm and}\qquad 
\frac{I_1(x)}{I_0^2(x)}e^{x\frac{I_1(x)}{I_0(x)}}=\frac{2\pi^2 x}{\eta}
\label{24}
\eea
A solution exists only for $\gamma^\prime<2\alpha(T)/\pi$. Else the density is uniform.
Thus the transition temperature from center symmetric (confining) to center-broken
(deconfining) occurs for $\alpha(T_c)/\gamma(T_c)=\pi/2N_c^3$ or 
$T_c^4=\frac{3}{8\pi}\frac{\Lambda^4}{\lambda^2}$ 
with $\lambda=g^2N_c/8\pi^2$.  For the dyon model, the fundamental string tension is given by
$\sigma_1=(N_c/\pi)\,{\rm sin}(\pi/N_c)\,\Lambda^2/\lambda$~\cite{DP}. Thus the model independent ratio
in $1+3$ dimensions

\be
\frac{T_c}{\sqrt{\sigma_1}}=\left(\frac {3\pi }{8N_c^2{\rm sin}^2(\pi/N_c)}\right)^{\frac 14}\rightarrow \left(\frac {3}{8\pi}\right)^{\frac 14}
\label{26}
\ee
%with the right-most value corresponding to the large $N_c$ limit. 
(\ref{26}) compares favorably 
to the lattice results~\cite{TEPER} even  for small $N_c$ as shown in Fig.~\ref{TC-data}.
At large $N_c$,  (\ref{26}) is consistent with the value of $\sqrt{3/2\pi}$ in the string model~\cite{ALVAREZ}.

\vskip 0.5cm
{\bf 10. Conclusions.\,\,}The hydrodynamical description of the Polyakov line captures aspects of the center
dynamics in Yang-Mills theory in terms of the gauge invariant density of eigenvalues. The hydro-static equations
yield solutions that interpolate between a center symmetric (confining) and a center-broken (de-confining) phase. The
transition temperatures normalized to the string tension compare well to the lattice results over a broad range
of $N_c$,  and asymptote the string model results at $N_c=\infty$.
The hydrodynamical set-up supports a hydrodynamical instanton  that describes 
the stochastic relaxation of the eigenvalues of the Polyakov line viewed as a fluid. The fluid supports sound waves
that can be used to estimate the probability of formation of Z(N$_c$)  bubbles. The relaxation of a fluid of holonomies
across the critical temperature may prove useful for understanding the onset of equilibration in a Yang-Mills plasma.

\vskip 0.5cm
{\bf Acknowledgements}
The work of YL and IZ is  supported in part  by the U.S. Department of Energy under Contracts No.
DE-FG-88ER40388. The work of PW is supported by the DEC-2011/02/A/ST1/00119 grant and 
the UMO-2013/08/T/ST2/00105 ETIUDA scholarship of the (Polish) National Centre of Science. 

%%%%%%%%%%%%%%%%%%%%%%%%%%%%%%%%%%%%%%%%%%%%%%%%%%%%%%%%%%%%%%%%%%%
 \vfil
\end{document}